\documentclass[aps,prl,twocolumn,superscriptaddress,showpacs,floatfix]{revtex4}

\usepackage{amsmath,amssymb,bm}
\usepackage{graphicx}

\begin{document}

\title{Origins of anomalous electronic structures of epitaxial graphene 
on silicon carbide}
\author{Seungchul Kim} 
\author{Jisoon Ihm}
\affiliation{Department of Physics and Astronomy, Seoul National University, Seoul 151-747, Korea}
\author{Hyoung Joon Choi}
\affiliation{Department of Physics and IPAP, Yonsei University, Seoul 120-749, Korea}
\author{Young-Woo Son}
\email[To whom correspondence should be addressed. E-mail:\ ]{youngwoo@konkuk.ac.kr}
\affiliation{Department of Physics, Konkuk University, Seoul 143-701, Korea}
\date{\today}
\begin{abstract}
On the basis of first-principles calculations, 
we report that a novel interfacial atomic structure occurs between 
graphene and the surface of silicon carbide, destroying the Dirac
point of graphene and opening a substantial energy gap there. 
In the calculated atomic structures, a quasi-periodic $6\times 6$ domain 
pattern emerges out of a larger commensurate $6\sqrt{3}\times6\sqrt{3}R30^\circ$ 
periodic interfacial reconstruction, resolving a long standing experimental 
controversy on the periodicity of the interfacial superstructures. 
Our theoretical energy spectrum shows a gap and midgap states at 
the Dirac point of graphene, which are in excellent agreement 
with the recently-observed anomalous angle-resolved photoemission spectra. 
Beyond solving unexplained issues in epitaxial graphene, 
our atomistic study may provide a
way to engineer the energy gaps of graphene on substrates. 
\end{abstract}
\maketitle
Graphene, a carbon allotrope, is a two-dimensional hexagonal network of carbon 
atoms which is formed by making strong triangular $\sigma$-bonds 
of the $sp^2$ hybridized orbitals~\cite{geim,neto}. 
The $\pi$-orbitals orthogonal to the hexagonal plane of graphene are responsible for its 
characteristic electronic properties, i.e., a relativistic dispersion relation near the Fermi 
level described by the massless free particle Dirac equation~\cite{geim,neto,novo1,zhang}. 
Following the report of successful fabrication of mechanically exfoliated graphene 
on the insulating SiO$_2$ surface~\cite{novo1,zhang}, 
tremendous efforts have been devoted to measure and exploit the novel physical 
properties of graphene~\cite{geim,neto}. 

On the other hand, it has been known for the last three decades that, when a wide 
band gap semiconductor silicon carbide (SiC) is heated up to 1300$^\circ$C, 
the monocrystalline graphite forms 
on the SiC (0001) face~\cite{starke,seyller,bommel,tsai,owman,forb,charrier,chen}. 
Now, by fine tuning the growth parameters, a 
single layer of graphene can be grown successfully on SiC~\cite{berger,heer}. 
These researches have stimulated interests in resolving fundamental material 
properties~\cite{starke,seyller} as well as 
applying the techniques to nanoelectronics~\cite{berger,heer}, 
with the merits of precise control of the number of 
layers of graphene~\cite{bost1,bost2,ohta} 
and a possible large scale production~\cite{heer}. 

Epitaxial graphene has demonstrated different 
physical properties compared to exfoliated graphene,
exhibiting many controversial experimental 
observations~\cite{chen,berger,heer,bost1,bost2,ohta,mallet,brar,emtsev,zhou}.
For example, scanning tunnelling microscopy (STM)
images~\cite{chen,berger,heer,mallet,brar} show a $6\times6$ hexagonal superstructure
with respect to the surface unitcell of 4H-SiC(0001) while 
low-energy electron diffraction (LEED) patterns indicate
a larger scale reconstruction with a 
$6\sqrt{3}\times6\sqrt{3}R30^\circ$ 
periodicity~\cite{starke,seyller,bommel,tsai,owman,forb,charrier,chen,berger,heer,mallet,emtsev}.
Moreover, the energy spectrum from STM~\cite{brar} and the angle-resolved photoemission 
spectroscopy (ARPES) measurements~\cite{zhou}  
show the energy gap which still defy precise 
interpretations~\cite{novo2}. 
Since the potential profiles induced by interfacial 
atoms will play a decisive role 
in the physical properties of graphene grown on SiC(0001)~\cite{geim,neto,zhou,novo2,gio}, 
it is indeed required to know the precise atomic 
geometries and the corresponding electronic structures of the system. 
However, in spite of many experimental observations, the atomic structures 
of the graphene (graphite)-SiC(0001) interfaces have not been uncovered 
yet except that they have a large scale reconstruction 
with the $6\sqrt{3}\times6\sqrt{3}R30^\circ$ 
periodicity~\cite{starke,seyller,bommel,tsai,owman,forb,charrier,chen,berger,heer,mallet,emtsev}.

In this Letter, we identify atomic and electronic structures of 
epitaxial graphene on 4H-SiC(0001) by large-scale first-principles calculations. 
In the relaxed atomic structures of the $6\sqrt{3}\times6\sqrt{3}R30^\circ$
periodicity with a single layer graphene, 
a quasi-periodic $6\times 6$ domain pattern appears, resolving 
the aforementioned disagreement between LEED patterns and the STM images.
The obtained novel domain pattern originates from 
interplay between strong bonding and lattice mismatch at the graphene-SiC(0001) interface.
With inclusion of another layer of graphene, 
the calculated electronic structures show a gap opening 
and midgap states at the Dirac point of graphene,
originating from sublattice symmetry breaking 
interactions between graphene and the interfacial superstructure in the system.
Simulated STM images and simulated ARPES spectra for the obtained
atomic structures show excellent agreements with 
several experimental data~\cite{chen,berger,heer,mallet,brar,emtsev,zhou}. 
Our study resolved the fundamental issues regarding the role of interfaces 
and substrates in altering physical properties of graphene, and thus
provides a way to control the energy gaps of graphene on substrates.

We study the atomic and electronic structures of graphene and interfacial carbons 
on 4H-SiC(0001) based on {\it ab initio} pseudopotential density functional methods~\cite{soler} within 
the local density approximation~\cite{alder} which are known to describe the structural and 
electronic properties of graphite quite well~\cite{matta,varchon}. 
The 4H-SiC(0001) substrate is modelled with four alternating silicon and carbon atomic layers, 
and one or two graphene layers are placed on top of the SiC substrate. 
The atoms belonging to the bottom layer of the slab are passivated by hydrogen.
To incorporate a large number of atoms in the system, we expand the wave function with localized
basis sets~\cite{basis}.
The basis sets and the pseudopotentials are thoroughly tested to reproduce 
the atomic and electronic structures of SiC, graphene and 
the  $\sqrt{3}\times\sqrt{3}R30^\circ$ model for 
epitaxial graphene studied in previous literatures~\cite{matta,varchon}
respectivley.
Based on the LEED measurements~\cite{starke,seyller,bommel,tsai,owman,forb,charrier,chen,berger,heer,mallet,emtsev}, 
the $6\sqrt{3}\times6\sqrt{3}R30^\circ$ periodicity is imposed to the SiC(0001) surface, 
which is equivalent to $13\times13$ times a graphene unit cell [Fig 1(a), (b)]. The atomic 
positions are determined by minimizing the total energy until the forces on each atom are 
less than 0.06 eV/\AA~while atoms belonging to the last two silicon and carbon layers are 
fixed to the bulk atomic structure of 4H-SiC.
To simulate ARPES spectra, 
wavefunctions of the $6\sqrt{3}\times6\sqrt{3}R30^\circ$ supercell obtained with a few k-points are
Fourier-transformed to the surface reciprocal space of graphene unit cell with dense k-points
and then integrated along the surface normal direction with attenuation corresponding 
to a photon mean free path of 5\AA. The calculated spectra are broadened by 30 meV in energy.

\begin{figure}[t]
\centering
\includegraphics[width=8.5cm]{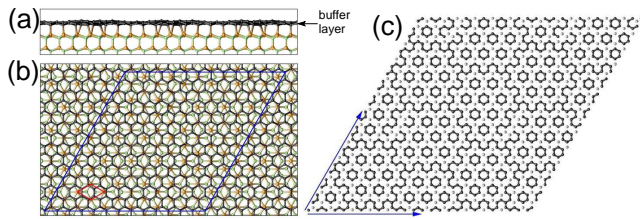}
\caption{(color online)
(a) Side and (b) top views of the atomic structure of the buffer layer with the 
$6\sqrt{3}\times6\sqrt{3}R30^\circ$ periodicity. 
The $6\sqrt{3}\times6\sqrt{3}R30^\circ$ supercell is denoted with blue lines and the 
$1\times 1$ surface unit cell of 4H-SiC(0001) with red. 
The carbon atoms in the buffer layer are denoted with black spheres 
and the silicon and the carbon atoms in the 4H-SiC with orange 
and green spheres, respectively. 
(c) Bonding characteristics of carbon atoms in the buffer 
layer. Carbon atoms with (without) $\sigma$-bonding to surface silicon atoms are represented with 
grey (black) dots. The $\pi$-bonds, represented by black lines, form a super-hexagonal pattern. 
Four times the $6\sqrt{3}\times6\sqrt{3}R30^\circ$ periodicity is drawn in order to display the super-hexagons 
clearly. The blue arrows are the unit vectors of the $6\sqrt{3}\times6\sqrt{3}R30^\circ$ supercell. 
}
\end{figure}

The obtained atomic structure, in the case of one layer of graphene on the SiC 
surface, displays a novel pattern of covalent bonding between carbon atoms in graphene 
and silicon atoms on the (0001) face [Fig. 1(c)]. 
We find that, due to the interplay of lattice mismatch and strong C-Si bonds, 
the $6\sqrt{3}\times6\sqrt{3}R30^\circ$ supercell is split into lattice matched 
regions, where carbon atoms are covalently bonded to surface silicon atoms with 
$\sqrt{3}\times\sqrt{3}R30^\circ$ periodicity with respect to the surface unitcell of SiC(0001). 
Outside the lattice matched regions, there are boundaries consisting of lines of carbon atoms (connected 
or disconnected with each other) without covalent bonding to surface silicon atoms. 
Our calculations show that the covalent bonding in the regions cannot be sustained over three or 
four units of the  reconstruction and that the carbon atoms with (without) 
covalent $\sigma$-bonding to silicon atoms move toward (away from) the SiC substrate. 
The small regions of covalent bondings follow the lattice symmetry of graphene so that all the lattice 
matched regions are shown to have large hexagonal shapes approximately [Fig. 1]. We find 
that the novel hexagonal pattern appearing in the present simulation is quite robust and does 
not depend on the details of calculations. 

\begin{figure}[t]
\centering
\includegraphics[width=8.5cm]{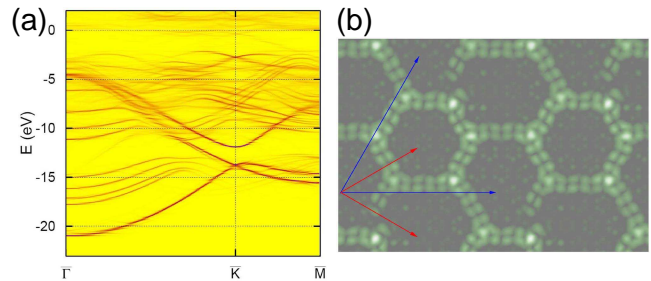}
\caption{(color online)
(a) Simulated spectrum for ARPES of 
the buffer layer on top of 4H-SiC(0001) in the graphene Brillouin zone. 
(b) Simulated STM image of the structure. 
Wavefunctions of which energies lie between 
the Fermi level ($E_F$) and 0.2 eV above $E_F$ are integrated 
and the image is taken in a plane located at 3 \AA~above the buffer layer. 
Bright (dark) regions correspond to a high (low) 
current in constant-height mode for STM. 
The $6\sqrt{3}\times6\sqrt{3}R30^\circ$ periodic lattice vectors are drawn in 
blue while $6\times6$ periodic ones in red.
}
\end{figure}

Our atomic structure for a single layer graphene on the SiC substrate [Fig. 1] is 
consistent with several experimental observations for the initial stage of graphene 
formation~\cite{bommel,tsai,owman,forb,charrier,chen,berger,heer,mallet,brar,emtsev}, 
in which the observed structure is called a carbon nanomesh~\cite{chen} or a 
buffer (or dead) layer~\cite{bost2,zhou,matta,varchon}. 
We will call it a buffer layer hereafter. First of all, our model for the buffer layer has 
the $6\sqrt{3}\times6\sqrt{3}R30^\circ$ reconstruction satisfying all the LEED 
measurements~\cite{starke,seyller,bommel,tsai,owman,forb,charrier,chen,berger,heer,mallet,emtsev}. 
Second, our simulated spectrum for the ARPES 
shows that $\sigma$-bands in the buffer layer appear clearly while
the linear $\pi$-bands around the Fermi energy 
are absent [Fig. 2(a)], compatible with the recent ARPES measurement~\cite{emtsev}. 
Instead, there are several flat bands above and below the Fermi energy
originating from $\pi$-orbitals of carbon atoms on the super-hexagonal boundaries. 
In our atomic structure [Fig. 1], the covalent bonding between graphene 
and the SiC surface breaks the 
hexagonal network of $\pi$-orbitals but preserves $\sigma$-bonds of $sp^2$ hybridization. 
Hence, the resulting band structure in the graphene Brillouin zone [Fig. 2(a)] 
shows no relativistic 
dispersion relation near the Fermi level of the system. 
Third, our simulated STM image shows an approximate $6\times6$ 
periodicity with respect to the surface unit cell of SiC(0001) [Fig. 2(b)]. 
This also bears a striking similarity to 
the observed STM images~\cite{chen,berger,heer,mallet,brar,varchon2}, 
although the $6\sqrt{3}\times6\sqrt{3}R30^\circ$ periodicity is imposed on our 
atomic structure. 
The bright regions in the simulated STM image [Fig. 2(b)] originate from 
the (broken) chains of $\pi$-orbitals with approximate super-hexagonal shapes in large scale 
and the dark regions correspond to domains of carbon atoms having strong $\sigma$-bonds to the 
surface silicon atoms.
Our atomic model for the buffer layer has a complete coverage of carbon
atoms on the SiC(0001) surface without any silion atoms on it. 
The model is compatible with not only the aforementioned spectroscopic measurement~\cite{emtsev} 
but also recent STM studies on the adsorption of metallic clusters~\cite{chen2} 
and fullerenes~\cite{chen3} on the buffer layer indicating clean carbon surfaces.

Next, we consider another graphene layer on top of the reconstructed buffer layer on 
4H-SiC(0001) surface [Fig. 3(a)]. The calculated atomic structure shows an almost free 
standing graphene with corrugations following the atomic structures underneath it [Fig. 3(a)]. 
The corrugation height is $\pm$0.17\AA, with the mean distance of 3.35 \AA~from the buffer layer. 
The simulated STM image [Fig. 3(b)] shows that all hexagonal networks of carbon atoms are 
clearly visible and approximate $6\times6$ periodic large hexagonal shapes are superimposed 
on it. This matches very well with the existing experimental data~\cite{heer,mallet,brar}, 
although the $6\sqrt{3}\times6\sqrt{3}R30^\circ$  periodicity is imposed on our atomic structure. 
In our calculations, the apparent $6\times6$ periodicity originates 
both from the atomic corrugations due to the 
underlying buffer layer and from the weak electronic interaction between graphene and the 
buffer layer. Thus, our results resolve the disagreement between the $6\times6$ periodicity from 
STM measurements and the $6\sqrt{3}\times6\sqrt{3}R30^\circ$ from the LEED measurements. 

\begin{figure}[t]
\centering
\includegraphics[width=8.5cm]{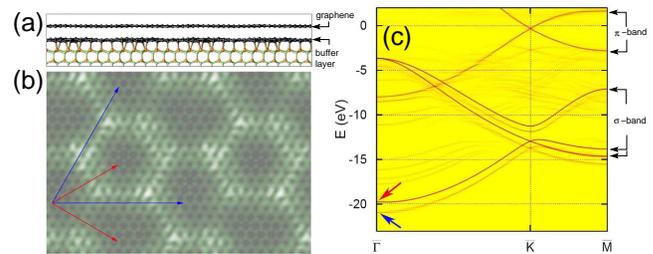}
\caption{(color online)
(a) Side view of the atomic structure of graphene on the buffer layer with the 
$6\sqrt{3}\times6\sqrt{3}R30^\circ$  periodicity. 
(b) Simulated STM image of graphene shown in (a). 
The unit vectors for the $6\sqrt{3}\times6\sqrt{3}R30^\circ$ periodicity are drawn in blue arrows 
and those for $6\times6$ in red. 
(c) Simulated ARPES spectrum for graphene shown in (a). 
The bottom of $\sigma$-bands in graphene is shown to 
be shifted up in energy (pointed by red arrow) compared with that in the buffer layer (blue 
arrow). The simulated STM image are obtained by the same method as in Fig. 2.
}
\end{figure}

We find that the simulated ARPES spectrum of graphene on top of the buffer layer 
show the characteristic $\pi$-bands of graphene as well as $\sigma$-bands [Fig. 3(c)]. 
The crossing point of two linear $\pi$-bands of graphene 
(called as the Dirac point~\cite{geim,neto}) 
is located slightly below the Fermi energy of the system. 
However, if looked closely, there is a gap opening at 
the Dirac point. We shall defer the discussion of the gap later. 
From Mullikan population analysis~\cite{soler}, 
graphene is found to be electron-doped with a density, $n\simeq 8.7\times10^{12}$/cm$^2$, 
consistent with experimental observations~\cite{bost1,bost2,ohta,emtsev,zhou}. 
It is also noticeable that the $\sigma$-bands of 
graphene are rigidly shifted up in energy compared with those of the buffer layer. This shift 
is observed in the recent ARPES measurement~\cite{emtsev} and the size of the shift (1.2 eV) is 
consistent with their observations~\cite{emtsev}. The energy shift of the $\sigma$-bands of graphene arises 
from a potential gradient due to a polar nature of the SiC surface. 

\begin{figure}[t]
\centering
\includegraphics[width=8.5cm]{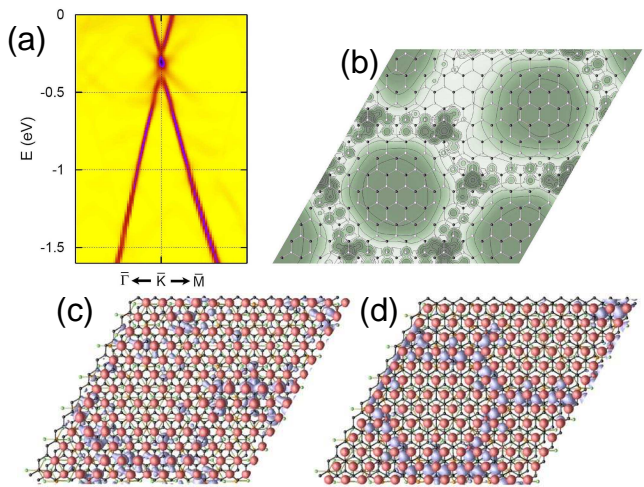}
\caption{(color online)
(a) The magnified view of the ARPES spectrum near the Dirac point. 
(b) Contour plot (contour spacing = 0.1 eV) for the 
potential generated by the buffer layer and SiC substrate only, 
drawn on a plane located at 3.35\AA~above the buffer layer. 
The bright (dark) color corresponds to high (low) potential. 
The hexagonal network (grey lines) for graphene is drawn to guide the eyes. 
The dark and bright dots represent two sublattices having
$\sim$140 meV averaged potential difference, respectively. 
(c) The squared amplitude of wavefunctions (isosurface of $3.0\times10^{-4}$/\AA$^3$) 
of which energy is located at the upper apex of the energy bands at K (shown in (a)). 
The amplitude of the wavefunction at graphene is denoted in red 
while the one at the buffer layer in light blue. 
(d) The squared amplitude of the wavefuntions whose energies are inside the energy gap at 
Dirac point (the averaged midgap state). The isosurface value is $3.0\times10^{-4}$/\AA$^3$. 
The red and blue colours for the amplitude follow the same scheme in (c).
}
\end{figure}

The electronic structure near the Dirac point shows a gap of 200 meV with the 
centre of the gap located at 320 meV below the Fermi energy [Fig. 4(a)]. It also shows 
midgap states inside the gap. The spectrum exhibiting the gap and midgap states is very 
similar with the recent ARPES observations~\cite{zhou}. The gap at the Dirac point in our 
calculation originates mainly from the interlayer coupling between graphene
and the buffer layer which breaks a sublattice symmtery in graphene~\cite{neto,zhou,novo2,gio}. 
Considering the valley and pseudo-spin symmetries exhibited 
in the electronic structure of graphene~\cite{geim,neto}, 
there are two possible ways of inducing a gap at Dirac points in a single layer of graphene. 
One is the mixing of electronic states with different pseudo-spins in the same valley and the other is the mixing 
of states belong to different valleys~\cite{neto,zhou,novo2,gio}. 
Due to the presence of the buffer layer, 
there exists a substantially different interaction
at atomic sites belonging to each sublattice of graphene [Fig. 4(b)], breaking the 
sublattice symmetry. The resulting wavefunctions near the Dirac point reflect such a broken 
symmetry so that the weight of the wavefunction on one sublattice is predominant over the 
other [Fig. 4(c)]. The intervalley mixing is found to have a negligible contribution to the gap. In case 
that the substrate is intentionally removed from the system in the calculation, the remaining 
graphene with present corrugations does not show any gap at the Dirac point. 
The midgap states originate from the interlayer coupling 
(the hopping energy of 0.2$\sim$0.3 eV between $\pi$-orbitals~\cite{neto}) 
between the $\pi$-states in graphene and the localized $\pi$-states 
on the boundaries between super-hexagons in the buffer layer [Fig. 4(d)].
The corresponding wavefunctions  
spatially spread out into graphene (the topmost surface 
of the system) [Fig. 4(d)] and thus can be detected by surface sensitive measurements such 
as ARPES. It explains the reason why the energy distribution curve shown in the recent 
ARPES measurement~\cite{zhou} has anomalous non-vanishing weights inside the gap at the Dirac 
point.

In summary, we have demonstrated that the interface between the SiC (0001) surface and 
graphene show the novel large scale atomic reconstruction and the fundamental electronic 
property of epitaxial graphene is altered due to the interface. The present atomistic study on 
the atomic and electronic structures of epitaxial graphene will play a crucial role not only in 
designing electronic circuits~\cite{heer} but also in explaining many other puzzling observations 
such as the absence of the quantum Hall effect and the weak 
Shubnikov-de Hass oscillations in the high mobility sample~\cite{heer,dara}. 

Y.-W. S. thanks H.-D. Kim and W. Chen for discussions. S. K and J. I acknowledge the support of 
the SRC program (Center for Nanotubes and Nanostructured Composites) of 
MOST/KOSEF.
H.J.C. acknowledges support from the KRF (KRF-2007-314-C00075) and from the 
KOSEF Grant No. R01-2007-000-20922-0. Y.-W. S. was supported by the KOSEF 
grant funded by the Korea government (MOST) No. R01-2007-000-10654-0. 
Computational resources have been provided by KISTI Supercomputing Center
(KSC-2007-S00-1011).


\begin{thebibliography}{99}
\bibitem{basis} We have used a single-$\zeta$ and a single-$\zeta$ plus
                polarization basis set for hydrogen and silicon atom respectively. 
                For carbon atom, a mixed basis set with a single and double-$\zeta$
                for $s$- and $p$-orbitals respectively, has been used.
\end{thebibliography}
\end{document}